# Generalized Integrated Functional Test for Regional Methylation Rates

Duchwan Ryu [1], Hongyan Xu [1], Varghese George [1], Shaoyong Su [2], Xiaoling Wang [2], Huidong Shi [3] and Robert H. Podolsky [4,*]

[1]Department of Biostatistics and Epidemiology, Georgia Regents University, Augusta, GA 30912, U.S.A.

[2]Georgia Prevention Institute, Department of Pediatrics, Georgia Regents University, Augusta, GA 30912, U.S.A.

[3]GRU Cancer Center, Department of Biochemistry and Molecular Biology, Georgia Regents University, Augusta, GA 30912, U.S.A.

[4]Department of Family Medicine and Public Health Sciences, Wayne State University, Detroit, MI 48201, U.S.A.

## ABSTRACT

**Motivation:** Methods are needed to test pre-defined genomic regions such as promoters for differential methylation in genome-wide association studies, where the number of samples is limited and the data have large amounts of measurement error.

**Results:** We developed a new statistical test, the generalized integrated functional test (GIFT), which tests for regional differences in methylation based on differences in the functional relationship between methylation percent and location of the CpG sites within a region. In this method, subject-specific functional profiles are first estimated, and the average profile within groups is compared between groups using an ANOVA-like test. Simulations and analyses of data obtained from patients with chronic lymphocytic leukemia indicate that GIFT has good statistical properties and is able to identify promising genomic regions. Further, GIFT is likely to work with multiple different types of experiments since different smoothing functions can be used to estimate the functional relationship between methylation percent and CpG site location.

**Availability and Implementation:** Matlab code for GIFT and sample data are available at http://biostat.gru.edu/~dryu/research.html.

Contact: rpodolsk@med.wayne.edu or dryu@gru.edu

## 1  INTRODUCTION

Genomic researchers are increasingly interested in identifying regions of DNA where methylation of the DNA has been altered in disease (Hansen et al., 2012; Hebestreit et al., 2013; Sofer et al., 2013; Sun et al., 2011; Wang et al., 2013). Most of these methods have been developed to identify differentially methylated regions (DMRs) where the regions are defined based on the pattern of differential methylation (Hansen et al., 2012; Hebestreit et al., 2013; Irizarry et al., 2009; Jaffe et al., 2012a; Jaffe et al., 2012b). The regions identified by many of these methods do not consider the location of the region relative to potentially important features such as promoters and CpG islands. Some of these existing methods are more general (e.g., Jaffe et al., 2012b), while some methods have been developed specifically for methods based on bisulfite sequencing (Hansen et al., 2012; Hebestreit et al., 2013).

The current methods tend to use functional data analysis methods (Ramsay and Dalzell, 1991; Ramsay and Silverman, 2010) where the functional relationship between methylation/differential methylation and location is modeled to estimate a subject-specific profile. These methods differ in how the subject-specific profile is smoothed, partly due to the extent to which the data already appear to exhibit a smooth relationship. Hansen et al. (Hansen et al., 2012) used a local-likelihood smoother which produces relatively smooth relationships, which was appropriate for the slowly changing methylation levels over a region observed in their data. Likewise, Hebestreit et al. (2013) used a triangular kernel that captured the step-like changes observed in their data. The data that we analyzed for this paper do not show as smooth of a function, and as such, we utilized wavelets to smooth the function.

Once a smoothed function is obtained, a test needs to be calculated for a given region. Hansen et al. (2012) used a test-statistic that was similar to a t-test for each CpG site, defined regions based on consecutive sites exceeding an arbitrary value for the test statistic, and used permutations for significance testing of the identified regions. Hebestreit et al. (2013) developed a method they called BiSeq, which used beta regression to test the significance of each CpG site, transform the resulting p-values into normalized z-scores, calculate the average normalized z-score for a given region, and compare the average normalized z-score to those obtained from resampling data. Other methods based on functional data analyses also exist (Coffey and Hinde, 2011; Faraway, 1997; Ramsay and Silverman, 2010; Yang et al., 2007).

Shen and Faraway (2004) provide an alternate approach that does not require smoothing provided the locations are fixed quantities, as can be assumed with genome-wide methylation data. Two models can be fit to each location, one in which all samples are pooled into one group and one in which the samples are kept in their respective groups. An F-test is then calculated where the residual sum of squares for each type of model is compared across all locations within a region.

In this paper, we estimate methylation profiles by estimating the functional relationship between $\beta$ and site location using wavelets in a manner similar to several papers by Morris *et al.* (Morris et al., 2006; Morris and Carroll, 2006; Morris et al., 2003). The wavelet method we use captures the spike-like features evident in the observed methylation profiles within a region for our data. We then use the fitted functions as the basis for developing a statistical test similar to that of Shen and Faraway (2004) to examine the differences between groups in a region. This method calculates an F-like statistic for a region based on the smoothed curves fit to a region by comparing the overall functional relationship to the average curve within each group.

The rest of the article is structured as follows. Section 2 develops the wavelet method for estimating the functional relationship for a given region. Section 3 develops the test procedure that we term the generalized integrated functional test (GIFT). Section 4 presents the results of a simulation study that examines the performance of GIFT and compares this test with the F-test of Shen and Faraway (2004). Section 5 applies our method to methylation data obtained by bisulfite sequencing in a study of chronic lymphocytic leukemia (CLL). We also compare our method with the F-test of Shen and Faraway (2004) and the method of Hebestreit et al. (2013). In doing so, we compare the use of different smoothing functions and a method that does not smooth the relationship.

## 2 WAVELET SMOOTHING TO ESTIMATE THE METHYLATION/LOCATION RELATIONSHIP

We focus here on the analysis of a set of CpG sites within a given genomic region (e.g., CpG island or promoter). Within a genomic region, we consider the percent DNA molecules that are methylated, $\beta$-value, to be a functional response in relation to the genomic location. We estimate the true profile function using wavelet denoising, a data-driven non-parametric regression technique. Although the exact genomic location of each CpG site is important, the exact functional relationship between $\beta$-value and genomic location does not have biological meaning. As such, we assume that CpG sites are equally spaced within a region, simplifying the analyses.

Wavelets are orthogonal families of basis functions that can represent other functions accurately and parsimoniously, as described in Vidakovic (1999). A wavelet series can approximate a continuous function $g(t)$ with $J$ scales such that

$$g(t) = \sum_{k=1}^{K_J} c_{J,k} \phi_{J,k}(t) + \sum_{j=1}^{J} \sum_{k=1}^{K_j} d_{j,k} \psi_{j,k}(t)$$

where $\phi_{j,k}(t)$ is a father wavelet basis function and $\psi_{j,k}(t)$ is a mother wavelet basis function for a location-scale decomposition of $g(t)$. The basis functions are dilations and translations of a father wavelet function and a mother wavelet function, e.g., $\phi_{j,k}(t) = 2^{-j/2}\phi(2^{-j}t-k)$ and $\psi_{j,k}(t) = 2^{-j/2}\psi(2^{-j}t-k)$. The coefficients are the wavelet coefficients. The smooth behavior of $g(t)$ at a coarse scale, $J$, is obtained using $c_{J,K_J}, \ldots, c_{J,1}$ as the smooth coefficients and $d_{j,K_j}, \ldots, d_{j,1}$, for $j = 1, \ldots, J$, as the detail coefficients to represent deviations of $g(t)$ at a finer scale. The coefficients can be obtained by taking the inner product of the function and the corresponding wavelet basis functions, i.e., $c_{j,k} = \int g(t)\phi_{j,k}(t)dt$ and $d_{j,k} = \int g(t)\psi_{j,k}(t)dt$. In practice, more efficient algorithms are available to calculate these coefficients. When $g(t)$ is defined on equally spaced $t = t_1, \ldots, t_m$, the wavelet coefficients are computed with the discrete wavelet transform (DWT) in just O(m) operations (Mallat and Hwang, 1992). Similarly, the inverse discrete wavelet transform (IDWT) may be used to project $c$ back into $g$ by multiplying the IDWT projection matrix $W$, i.e., $g = cW$, where $W^T$ is the transpose of the DWT orthogonal projection matrix.

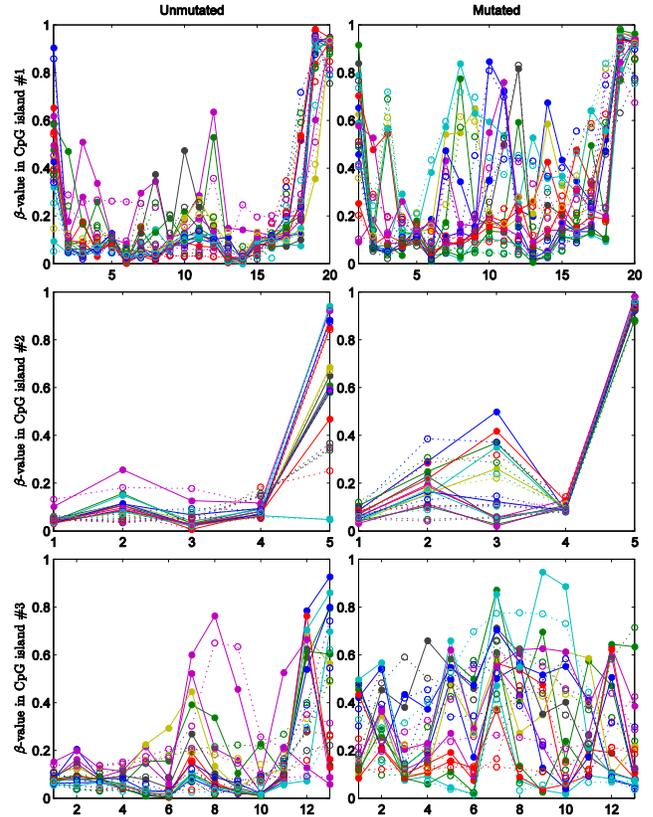

**Fig. 1.** Examples of wavelet de-noised profiles of methylation rates. In each plot, the observed methylation rates are shown as connected dotted lines, and de-noised curves are shown as solid-lines for the corresponding color.

Each subject's methylation profile is estimated as follows. Let $\beta_i = [\beta i\ (t1\ ), \ldots, \beta i\ (tm)]$ denote a profile of observed methylation rates from the sample $i$ on $m$ CpG sites. Because the methylation rate ranges from 0 to 1, we use logit-transformed methylation rates, $\beta_i^* = \log\left(\dfrac{\beta_i + \alpha}{1 - \beta_i + \alpha}\right)$, where $\alpha$ is a small adjustment factor to prevent a zero denominator. In this paper, we set $\alpha = 10^{-4}$. The following procedure is used for de-nosing the $\beta_i^*$ profile: (1) Use DWT to Project $\beta_i^*$ into the wavelet domain and obtain the empirical wavelet coefficients $c^*$; (2) Set any coefficients less than a specified threshold to 0; (3) Use IDWT to project the thresholded coefficients back to the original data domain and establish the de-noised functional values $g_i^*$; and (4) rescale $g_i^*$ as

$$g_i = \left\{(\alpha+1)\exp(g^*) - \alpha\right\} / \left\{1 + \exp(g^*)\right\}.$$

By taking a small number of dominant wavelet coefficients, this procedure removes the white noise distributed equally in all wavelet coefficients (Daubechies, 1992; Morris et al., 2006). We use the Matlab® Wavelet Toolbox with Daubechies's wavelet to fit the wavelet coefficients, and determine the threshold using the Birgé-Massart (1997) strategy. Fig. 1 illustrates the de-noised methylation profile functions in three exemplary CpG islands in a study comparing cancer specimens that differ in disease aggressiveness.

These de-noised functions from individual samples are used as the basis for a statistical test we develop to compare groups of methylation profiles in the next section.

# 3 GENERALIZED INTEGRATED FUNCTIONAL TEST

We consider the methylation profile $\boldsymbol{\beta_{ij}} = [\beta_{ij}(t_1),...,\beta_{ij}(t_m)]$ observed in a genomic region from the sample $j$ in the group $i$, where the genomic region includes CpG sites located between $a$ and $b$, i.e., $t_l \in [a,b]$, $l = 1,...,t_m$. The observed profile is assumed to follow a true function with noise. Suppose that we estimate the true function through the wavelet de-noising method of Section 2, and let $g(t)$ denote the estimated true function for the sample. Let $\bar{g}_i(t) = 1/n_i \sum_{j=1}^{n_i} g_{ij}(t)$ denote the average function within group $i$ and let $\bar{\bar{g}}(t) = 1/k \sum_{i=1}^{k} \bar{g}_i(t)$ denote the overall average function across all groups (Fig 2a.). This method contrasts with the method of Shen and Faraway (2004) in which a logistic regression is fit to each CpG site (Fig. 2b).

To compare functions $g_{ij}$ from $k$ groups, i.e., i = 1, . . . , $k$, Ramsay et al. (2009) suggested a functional F-test statistic, at a given location $t$, such that

$$F(t) = \frac{Var\{\mathbf{g}(t)\}}{\frac{1}{k}\sum_{i=1}^{k}\frac{1}{n_i}\sum_{j=1}^{n_i}\{g_{ij}(t) - \bar{g}(t)\}^2}$$

where $\bar{\mathbf{g}}(t) = [\bar{g}_1(t),...,\bar{g}_k(t)]$ are values of mean functions over $k$ groups evaluated at $t$. The functional F-test describes the group difference over the region $t \in [a,b]$ and identifies locations where the groups are significantly different point-wisely. However, this approach does not provide a single-test to determine whether the groups have significantly different functions in the region as a whole.

We consider an integrated version of the test statistic over the given region to summarize the evidence for an overall difference within the region. Like the sums of squares in the usual F-test, we define two quantities: the integrated sum of squares of treatment (ISST) and the integrated sum of squares of error (ISSE), over the range of $t$ such that

$$ISST = \frac{1}{k}\sum_{i=1}^{k}\int_a^b \{\bar{g}_i(t) - \bar{\bar{g}}(t)\}^2 \, dt,$$

$$ISSE = \frac{1}{k}\sum_{i=1}^{k}\frac{1}{n_i}\sum_{j=1}^{n_i}\{g_{ij}(t) - \bar{g}_i(t)\}^2 \, dt.$$

To examine the functional difference of groups, we construct a generalized integrated functional test (GIFT) statistic based on ISST and ISSE:

$$GIFT = \frac{ISST}{ISSE}.$$

We determine the p-value for GIFT by nonparametric permutations rather than assuming a parametric distribution of the test statistic. By randomly shuffling the functions $g_{ij}(t)$ with respect to the group index $i$, we can investigate the probability that the difference among groups is larger than the current difference.

# 4 SIMULATION STUDY

We used simulation to compare the performance of GIFT with the F-test suggested by Shen and Faraway (2004). We simulated regions with two groups of methylation functions. Functions for each subject in each group were generated by adding variability to the corresponding group function, and methylation rates were generated by adding white noise to these subject-specific functions. We considered profiles of methylation rates over 10 locations from 6 individuals for each group.

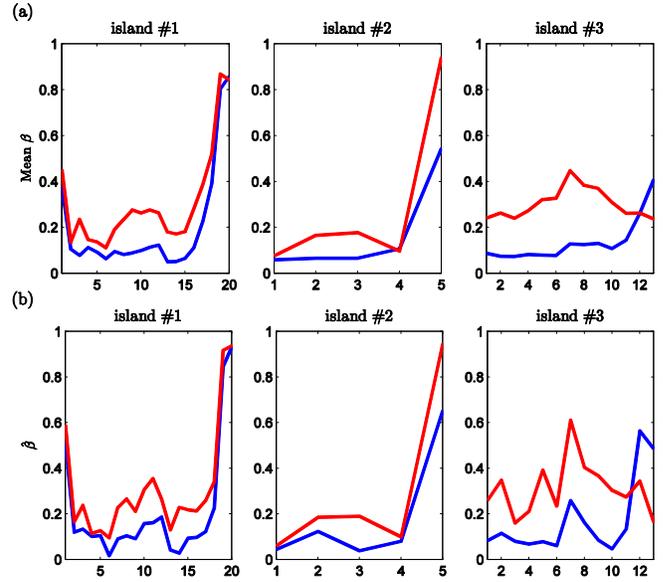

**Fig. 2.** Group-specific average de-noised profiles. Blue lines denote the curves for the unmutated group and the red lines denote the mutated groups. (a) Average methylation functions achieved by wavelet de-noising method. (b) Fitted curves driven by logistic regressions for each CpG site.

We simulated four genomic regions with different functional patterns under two effect sizes. Let $\beta_{ij}(t)$ denote the simulated profile of methylation rates and let $g_i(t)$ denote the true methylation function from group $i = 1, 2$ and the sample $j = 1, . . . , 6$, for the location $t = 1, . . . , 10$. The functions for the genomic regions for the two groups for simulations with large differences between groups were as follows:

Genomic Region 1,

$g_i(t) = \left[1 + \exp\left\{\mu_i + \delta_{0j} + \delta_{1j}\frac{1}{3}(2t-11)\right\}\right]^{-1}$, with $\mu_1 = \log(4)$ and $\mu_2 = $ -log(4);

Genomic Region 2,

$g_i(t) = \left[1 + \exp\left\{\delta_{0j} + (\delta_{1j} - \mu_i)\frac{1}{3}(2t-11)\right\}\right]^{-1}$, with $\mu_1 = 1$ and $\mu_2 = -1$;

Genomic Region 3,

$g_i(t) = \left[1 + \exp\left\{\mu_i + \nu_i|\frac{2}{9}(2t-11)|^{\frac{1}{2}} + \delta_{0j} + \delta_{1j}\frac{2}{9}(2t-11)\right\}\right]^{-1}$, with $\mu_1 = 2.5$, $\nu_1 = $ -2, $\mu_2 = $ -2.5, $\nu_2 = 2$,

where $\delta_{0j}$ and $\delta_{1j}$ are random coefficients to generate an individual-specific function and with $\delta_{0j} \sim N(0, 0.5^2)$ and $\delta_{1j} \sim N(0, 0.1^2)$. The observed methylation rates were obtained by adding independent Gaussian errors, $\varepsilon \sim N(0, 0.5^2)$ to the subject-specific function at each location (Fig 3).

The fourth genomic region was simulated as functions having three peaks:

$$g_{it}(t) = \begin{cases} -\frac{1}{5}\left(c_{i1j} + c_{i3j} + \mu_{i1j}\,|t-2|^{0.7}\right), & \text{if } t < 3.5 \\ -\frac{1}{5}\left(c_{i3j} + \mu_{i2j}\,|t-5|^{0.5}\right), & \text{if } 3.5 \le t < 6.5 \\ -\frac{1}{5}\left(c_{i2j} + c_{i3j} + \mu_{i3j}\,|t-8|^{0.7}\right), & \text{if } t \ge 6.5 \end{cases}$$

where $\mu_{ikj}$ are random coefficients for group $i$ ($i = 1,2$) from uniform distributions such that $\mu_{11j}, \mu_{13j} \sim U(0.2, 0.8)$, $\mu_{21j}, \mu_{23j} \sim U(1,3)$, $\mu_{12j} \sim U(0, 0.4)$, $\mu_{22j} \sim U(0,1)$. The coefficients $c_{i1j}$ and $c_{i2j}$ are used to create three peaks, and the coefficients $c_{i3j}$ keep the minimum values of functions as 0.06. These mean functions were then used to define the methylation proportion,

$$\beta_{ij}(t) \sim Beta\left[10 g_{ij}(t), 10\{1 - g_{ij}(t)\}\right] \text{ (Fig 3)}.$$

Similar group functions were used for all four regions with smaller differences between groups. We also simulated the four genomic regions in which the groups did not differ in the mean methylation function. Each region and effect size was simulated 1000 times.

The simulated profiles were analyzed in two ways. We first used wavelet smoothing to estimate the subject-specific functions, and then we used GIFT to test for differences between groups. P-values for GIFT were determined based on permutations. We also tested for differences between groups using two logistic regression models fit for each site: a null model, ignoring group when fitting the model; a full model, fitting a logistic regression model separately for both groups. Based on these two types of models, we calculated the region F-test as described by Shen and Faraway (2004):

$$F = \frac{(rss_{null} - rss_{full})/(p - q)}{rss_{full}/(n - p)}$$

where $rss$ are residual sums of squares from each model, $p$ and $q$ are the number of parameters used for each respective model, and $n$ is the total number of subjects. The degrees of freedom for this F-test were calculated as $df_1 = \lambda(p - q)$ and $df_2 = \lambda(n - p)$, where $\lambda$ is an adjustment factor, $\lambda = trace(E)^2/trace(E^2)$, and E is the empirical covariance matrix.

The p-value distribution under the null hypothesis should be close to uniform. Such a distribution was observed for GIFT for all four simulated genomic regions, while the p-value distribution for the F-test had a mode at approximately 0.5 (Supplemental Fig. 1). These results suggest that GIFT controls type 1 error better than the F-test.

The p-value distribution under the alternative hypothesis should be shifted toward small p-values, which was observed for both GIFT and the F-test. Although both GIFT and F-test had the expected p-value distribution, the F-test had smaller average p-values when a large difference in mean curves was simulated, regardless of the genomic region (Table 1, Supplemental Fig 2). However, the GIFT had slightly better power when the effect size was smaller as reflected by the distribution of p-values (Supple-mental Fig 2). These results are likely reflective of the F-test being conservative under the null hypothesis, and the power of this test overcoming the conservative nature once effect sizes are sufficiently large. Effect sizes in biological experiments tend to be small, especially in human studies. As such, we expect the GIFT to be a more powerful test in most experimental situations.

**Table 1.** Average p-values for simulated methylation regions.

| CpG | Small Effect Size | | Large Effect Size | |
|---|---|---|---|---|
| Region | GIFT | F-test | GIFT | F-test |
| 1 | 0.0371 | 0.0407 | 1.1e-3 | 7.0e-7 |
| 2 | 0.0481 | 0.0891 | 9.9e-4 | 5.9e-8 |
| 3 | 0.0540 | 0.0608 | 1.1e-2 | 6.0e-4 |
| 4 | 0.0506 | 0.0636 | 2.3e-2 | 1.4e-2 |

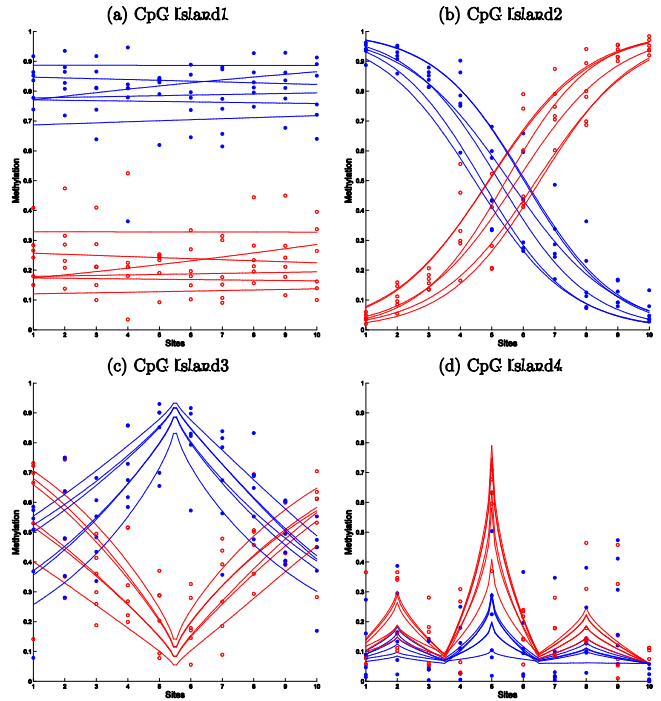

**Fig. 3.** Examples of true individual functions $g_{ij}(t)$ and simulated profiles of methylations $\beta_{ij}(t)$ with obvious different patterns. In each plot, lines describe true individual methylation functions generated by adding variations to corresponding group methylation functions, and dots and circles are methylation profiles based on individual true methylation functions.

## 5 ANALYSIS OF REGIONAL DIFFERENTIAL METHYLATION IN CHRONIC LYMPHOCYTIC LEUKEMIA DATA

### 5.1 Chronic Lymphocytic Leukemia and Transcription Factor Binding Site Data

We used data from a study of chronic lymphocytic leukemia (CLL), a B-cell lymphoma mainly of adults that is a very heterogeneous disease. Mutations within Ig VH genes are known to be associated with the aggressiveness of CLL; CLL patients with an unmutated Ig VH gene usually have a poorer prognosis (Damle et al., 1999; Hamblin et al., 1999). CD38 levels are known to be associated with Ig VH mutation status (Damle et al., 1999) and prognosis (Hamblin et al., 1999), with patients having lower levels progressing more slowly.

Reduced representation bisulfite sequencing (RRBS) (Meissner et al., 2005) was used to measure methylation levels in 11 CLL samples (Pei et al., 2012). The RRBS technology provides estimates of the percent of DNA molecules that are methylated, $\beta$, for any CpG site that was sequenced with a typical run providing $\beta$ estimates for approximately 2 million CpG sites. The samples were categorized as low- vs. high-risk based on CD38 levels, with seven samples having low CD38 levels (CD38<=30; low-risk) and four samples having high CD38 levels (high-risk). The RRBS data that we analyze have already been cleaned and aligned as described in Pei et al. (2012).

Changes in methylation are generally expected to result in changes in transcription. We decided to use ENCODE data to define candidate promoter regions (Rosenbloom et al., 2013) since dynamic methylation tends to be more distal at enhancers and other transcription factor binding sites (Ziller et al., 2013). We defined candidate transcription factor binding regions using ChIP-seq data (experimental score $\geq 500$) for three potentially relevant lymphoblastoid, B-lymphocyte cell lines: GM12878, GM12891, GM12892. CpG sites within regions were filtered to include only those sites with data for all subjects. Further, we filtered sites based on the total number of reads for each subject since we wanted to investigate the extent to which read depth for a CpG site influences results. In one set of analyses, only sites having at least 5 total reads were included, while in another set of analyses we only included those sites with at least 20 total reads. We focused our analyses on sites on two randomly selected chromosomes, 7 and 19. The analyses that contained only sites with at least 5 total reads for each subject included 574 regions with an average of 9.8 sites for each region. The analyses that contained only sites with at least 20 total reads for each subject included 324 regions also with an average of 9.8 sites for each region.

We annotated each transcription factor binding site for gene information using the hg18 build of the UCSC Genome Browser. We recorded any genes overlapped by the specific candidate transcription factor binding region, including genes in which the region was wholly contained within a gene. For regions with not overlapping genes, we continued searching for genes near each region beginning $\leq 500$ bases and expanding beyond this promoter region if no gene was found.

## 5.2 Methods Used for Comparison

We compared several methods: (1) wavelet de-noising and GIFT (Wavelet/GIFT); (2) the F-test we used in section 4; (3) BiSeq (Hebestreit et al., 2013) with default settings [BiSeq (h=80)]; (4) BiSeq with less smoothing [BiSeq (h=30)]; (5) smoothed methylation percents obtained from BiSeq using default settings and analyzed using GIFT [BiSeq (h=80)/GIFT]; and (6) smoothed methylation percents obtained from BiSeq with less smoothing and analyzed using GIFT [BiSeq (h=30)/GIFT]. For BiSeq testing, we used only those functions in the BiSeq package needed to test for differential methylation of a given region. In doing so, we did not cluster the CpG sites into candidate regions, nor did we prune regions following regional significance tests. These analyses allow us to both compare methods for testing for regional differences in methylation as well as evaluate the effect of different smoothing methods.

The GIFT and F-test both produce p-values for each region, while BiSeq testing produces only q-values (Storey, 2003). Q-values were also calculated using the p-values for all GIFT and F-test results using the qvalue package in R (Dabney et al., 2014).

## 5.3 Comparison Results

We first evaluated the methods under the null hypothesis by analyzing one random permutation of the data and examining the distribution of p-values. The null hypothesis should hold under one random permutation of the data, and as such, the distribution of the p-values should be uniform. BiSeq testing was not included in this comparison because BiSeq does not readily produce regional p-values.

Wavelet/GIFT and BiSeq (h=30)/GIFT both produced the expected uniform distribution of p-values regardless of whether the methylation data were filtered to have at least 5 or at least 20 total reads per site for each subject (Fig. 4). However, BiSeq (h=80)/GIFT only produced a uniform distribution of p-values for the 5-read filtered data. The F-test never resulted in a uniform distribution of p-values. These results suggest that both wavelet smoothing and rougher smoothing in BiSeq combined with GIFT maintain the correct type 1 error, at least for the CLL data.

Most methods used to analyze the observed CLL methylation data produced p-value distributions skewed towards small p-vaues, as expected (Fig. 5). However, the p-value distributions for the F-test were convex in contrast to the expected concave shape, indicating that assumptions for the F-test must have been violated. The stronger skew observed for BiSeq/GIFT suggests that these smoothing methods have greater power than using wavelet smoothing with GIFT.

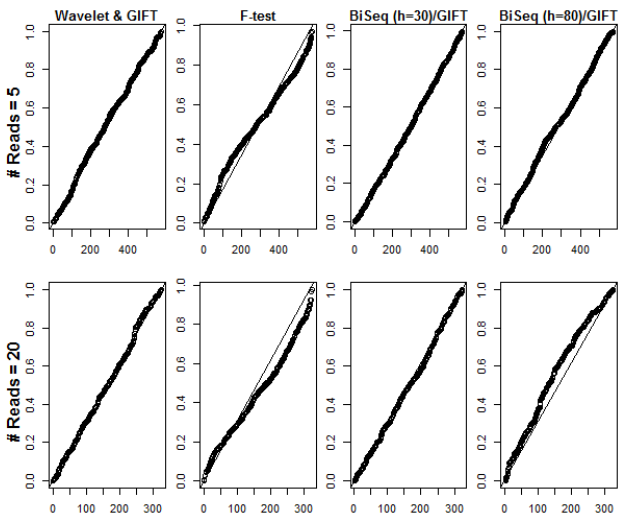

**Fig. 4.** Uniform probability plots of the observed p-values from analyses of one random permutation of the CLL data.

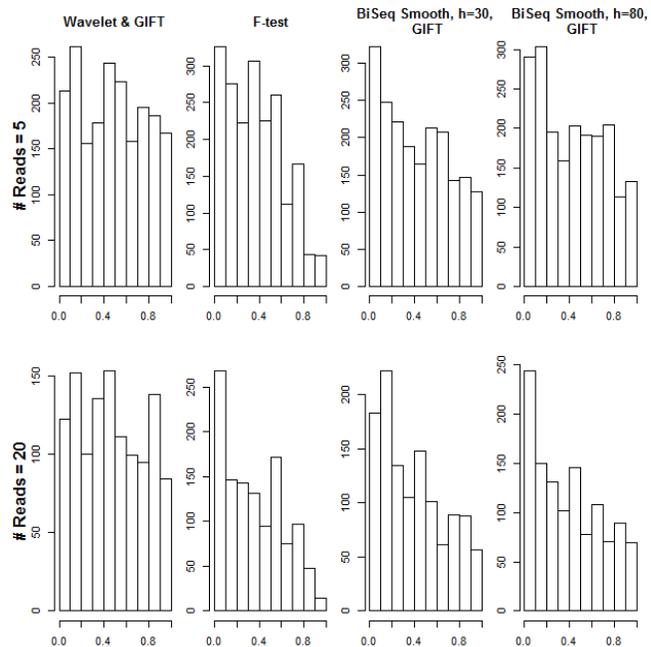

**Fig. 5.** P-value distributions based on analyses of observed CLL methylation data.

**Table 2.** Minimum p-value and q-values for analyses of the observed CLL data.

| | 5-Read | | 20-Read | |
|---|---|---|---|---|
| Analysis | p-value | q-value | p-value | q-value |
| Wavelet/GIFT | 0.0004 | 0.2224 | 0.0030 | 0.4375 |
| F-test | 0.0009 | 0.0435 | 0.0002 | 0.0024 |
| BiSeq (h=30)/ GIFT | 0.0004 | 0.1384 | 0.0012 | 0.1349 |
| BiSeq (h=80)/ GIFT | 0.0024 | 0.4102 | 0.0004 | 0.0647 |
| BiSeq (h=30) | -- | 0.05 | -- | 0.20 |
| BiSeq (h=80) | -- | 0.10 | -- | 0.01 |

Q-values were used to adjust for multiple testing. Four of the 5-read regions had q≤0.05 based on the F-test, while 88 of the 20-read regions were significant at this level. The q-values for the F-test are not likely accurate, however, due to the shape of the p-value distribution. Importantly, the largest q-value for the 5-read analyses was 0.10 and the largest q-value for the 20-read analyses was 0.08.

Two of the 5-read regions had q≤0.05 based on BiSeq (h=30). No region had q≤0.05 for the same analysis when the default smoothing (h=80) was used. In contrast, one region had q≤0.05 for the 20-read analyses based on BiSeq (h=80). None of the GIFT analyses yielded q≤0.05 (Table 2).

Although GIFT did not identify any regions with q≤0.05, the q-value for the smallest p-value was not always very large (Table 2). Based on these observed minimum p- and q-values, we decided to call any region with p≤0.0005 significant for tests producing a p-value. We also called all regions significant for BiSeq testing using q≤0.25, the approximate smallest q-value observed for wavelet/GIFT.

Using these criteria, 19 5-read and 9 20-read regions were significant. Eight of the 19 significant 5-read regions have some connection to CLL based on a literature search (Supplemental Table 1). Four of the 9 significant 20-read regions also had some connection to CLL (Supplemental Table 3).

Both of the 5-read regions with q≤0.05 using BiSeq (h=30) were also significant with BiSeq (h=80; q=0.1 and q=0.2; Supplemental Tables 1 & 2). The first of these regions was near a hypothetical protein, while the second region was close to two genes, one downstream from the region (carnitine palmitoyltransferase 1C isoform 2, CPT1C) and one upstream (HMT1 hnRNP methyltransferase-like 2, PRMT1). We are unaware of any potential function for CPT1C that would be implicated in CLL. On the other hand, PRMT1 is an arginine N-methyltransferase and increased expression of this gene may play a role in many types of cancer (http://www.ncbi.nlm.nih.gov/gene/3276). This gene is involved in B-cell receptor silencing (Yang and Reth, 2010), and the protein disrupts transcriptional repression of RUNX1 which is known to play a major role in B-cell differentiation (Tjchon et al., 2013).

Four of the 5-read regions were significant using wavelet/GIFT, but were not significant with the other methods. Only one of these four regions was close to a gene with potential connection to CLL (MIDN). This gene has been observed to be differentially expressed in control-transfected and CD5-transfected B cells (Gary-Gouy et al., 2007).

One of the 5-read regions detected by BiSeq (h=30)/GIFT was close to a gene with a connection to CLL: ZNF566. The number of single nucleotide mutations in ZNF566 were observed to increase in one patient during disease progression and following relapse after treatment with ofatumunab (Schuh et al., 2012). This region was also significant for all BiSeq-smoothing methods.

Two other 5-read regions with nearby CLL genes were significantly only with BiSeq testing: PCOLCE and SPIB. PCOLCE codes for a precursor for a fibrillar collagen, and such collagens are known to be involved in altered lymphocyte tracking and adhesion in CLL (Mayr et al., 2005; Mikaelsson et al., 2005; Zheng et al., 2002). Expression of SPIB has been shown to decrease during normal B cell differentiation (Rui et al., 2011).

The single 20-read region with q≤0.05 was significant using BiSeq/GIFT regardless of the degree of smoothing used (Supplemental Tables 3 & 4). This region was located within an intron of GNG7 (guanine nucleotide binding protein, a G-protein), and the expression of this gene is altered in CLL (Seifert et al., 2012). Similar to the results for the 5-read analyses, regions near SPIB and PRMT1 were significant with at least one of the methods used. One region found was significant only in the 20-read analyses and close to a gene with potential importance to CLL, CNN2. Patients with a 13q14.3 deletion had decreased expression of CNN2 and this deletion is common in CLL (Mian et al., 2012). Overall, these results suggest that our analyses have identified strong candidate regions with altered methylation signatures, although no single method identified all regions.

# 6 DISCUSSION AND CONCLUSION

GIFT is a general method for using functional data analyses methods to test for regional differential methylation. The p-value distributions generated by analyses of the permuted CLL data suggest that analyses with either wavelet smoothing or the rougher BiSeq smoothing better control type I error. Our data show a relatively "rough" relationship between $\beta$ and CpG site location, and the results favoring the rougher smoothing methods are likely a result of this observed relationship. While the CLL data did exhibit such a "rough" relationship, other studies have not shown as rough a relationship (Hansen et al., 2012; Hebestreit et al., 2013; Tavolaro et al., 2010). Our results point to the need to choose the appropriate smoothing technique in analyzing regional methylation data.

One limitation of wavelet smoothing is the inability to handle regions with missing data. The results presented herein were therefore limited to only regions with no missing data. As such, it is difficult to evaluate how GIFT would perform in general in the presence of missing data. Future studies should focus on the use of such functional tests in the presence of missing data. The positive results obtained using BiSeq smoothing indicate that these smoothing methods work in many situations. Further, BiSeq smoothing is able to estimate the functional relationship between $\beta$ and CpG site location taking into account missing data.

Unfortunately, our results do not identify a single method (GIFT with differing smoothing functions or BiSeq testing) as being best with the CLL data. Likewise, it is likely that no single method will work with every study. The various methods make different assumptions with regard to the expected relationship between $\beta$ and CpG site location, as well as with regard to the basis for the statistical test. It is possible that the ideal analysis of genome-wide methylation data will require the use of multiple methods in a manner similar to the analyses of gene expression data for determining whether sets of genes are differentially expressed (Emmert-Streib and Glazko, 2011; Glazko and Emmert-Streib, 2009).

One challenge in the analysis of genome-wide methylation data is the fact that changes in methylation have context-specific effects (Irvine et al., 2002). The regions that we identified as being poten-tially important were not necessarily located in the promoter region of genes or in CpG islands. Methylation differences outside of CpG islands are more likely to be dynamic (Jones, 2012) suggesting that the regions we have identified may be more dynamic. Still, our findings may be related to the way in which candidate regions were defined. The other methods that have been developed to identify differentially methylated regions focus on identifying the region based on the methylation data itself. Our approach was to use other genomic data to identify the candidate region, and hence we used GIFT which was designed to test for differential methylation of a region specified from other data. Ideally, multiple types of genomic data should be jointly analyzed, identifying the region using non-methylation data and methods such as those of Kheradpour and Kellis (Kheradpour and Kellis, 2014), and testing for differential methylation based on the methylation data. Such integrated analyses deserve more attention.